\documentclass[conference]{IEEEtran}
\usepackage{times}
\def\ninept{\def\baselinestretch{.95}\let\normalsize\small\normalsize}

\usepackage{url}
\let\labelindent\relax
\usepackage{enumitem}
\usepackage{array}
\setitemize[0]{leftmargin=*,partopsep=1pt,topsep=1pt}
\usepackage{verbatim}
\ifCLASSINFOpdf
   \usepackage[pdftex]{graphicx}
    \usepackage{subfig}

\else

\fi

\usepackage{times}

\begin{document}
%
\title{MetaPar: Metagenomic Sequence Assembly via Iterative Reclassification}

\author{\IEEEauthorblockN{Minji Kim$^{*}$, Jonathan G. Ligo$^{*}$, Amin Emad, Farzad Farnoud (Hassanzadeh)\\Olgica Milenkovic and Venugopal V. Veeravalli\\
$^{*}$The first two authors contributed equally to this paper.}
\IEEEauthorblockA{Department of Electrical and Computer Engineering, University of Illinois, Urbana-Champaign\\
Email: \{mkim158,ligo2,emad2,hassanz1,milenkov,vvv\} [at] illinois.edu}}

\maketitle
\begin{abstract}
We introduce a parallel algorithmic architecture for metagenomic sequence assembly, termed \emph{MetaPar}, which allows for significant reductions in assembly time and consequently enables the processing of large 
genomic datasets on computers with low memory usage. The gist of the approach is to iteratively perform read (re)classification based on phylogenetic marker genes
and assembler outputs generated from random subsets of metagenomic reads. Once a sufficiently accurate classification within genera is performed, de novo metagenomic assemblers (such as Velvet or IDBA-UD) or reference based assemblers may be used for contig construction. We analyze the performance of \emph{MetaPar} on synthetic data consisting of 15 randomly chosen species from the NCBI database~\cite{wp} through the effective gap and effective coverage metrics.
\end{abstract}

\IEEEpeerreviewmaketitle
\ninept
\section{Introduction} \label{sec:intro}
Metagenomics is a scientific discipline devoted to the study of complex microbial samples in the environment. Unlike classical genomics, where one is faced with the task of processing samples corresponding to one organism, metagenomics is concerned with samples that consist of a mixture of genetic material of different species and strains of bacteria or viruses~\cite{R2004} within a host. With raw data file sizes capable of exceeding hundreds of gigabytes, metagenomic data poses significant new challenges in Big Data signal processing and analysis since genomic assembly is computationally difficult even for single species analysis~\cite{R2004,Q2010,Pengetal2012}. 

Although significant progress was made in the last few years on developing new methods for metagenome assembly~\cite{Zerbino2008,Li2010,Pengetal2012,W2012}, the problem of accurate metasequence profiling  remains wide open. To hasten the progress of this new discipline, InnoCentive recently launched a special competition under the auspices of the U.S. Defense Threat Reduction Agency (DTRA) in metagenomic de novo assembly. Specialized access to powerful computers and clusters were given to all active participants for processing relatively small amounts of data (tens of gigabytes). In the absence of such strong computational support, large metagenomic assembly appears to be infeasible. One way to mitigate this issue is to break down the metagenomic data into smaller subsets that may be processed independently and in parallel, using modest computational power. This is the gist of the assembly approach \emph{MetaPar} for parallel metagenomic assembly that we proceed to describe in the remainder of the paper.

\emph{MetaPar} mitigates the need for computationally demanding full metagenomic assembly by using an iterative two-stage read classification technique. In the first stage, the microbial identification tool \emph{MetaPhyler}~\cite{Liuetal2010}  is used for providing a rough profile of organisms present in the metagenomic mix. By aligning the reads to \emph{all genomes} of organisms within the identified genera via \emph{Bowtie2}~\cite{LS2012}, one obtains a rough partition of the reads into subgroups. Reads within different identified subgroup are assembled in parallel, producing contigs that may be run through BLAST (Basic Local Alignment Search Tool)~\cite{A1990} to verify the accuracy of the classifier. After this first classification step, some reads may remain unaligned, and require alternative means of processing. 

Two options may be pursued in the second round of classification, depending on the number of unaligned reads. If the number of reads is prohibitively large so as not  to allow one-pass assembly with a standard metagenomic assembler, the reads are \emph{randomly} partitioned into subgroups small enough to be assembled. All assemblies are performed in parallel. Unaligned reads are iteratively reassigned between assemblers until no changes in the assembled contigs are reported or until a maximum number of iterations is reached. On the other hand, if the number of reads is small enough to allow for one pass assembly, the same procedure as outlined for the initial step is performed. Related ideas involving dynamic classification of reads were described in~\cite{W2012}, but for the purpose of \emph{single genome} assembly. \emph{MetaPar} is a simple parallel (and distributable) metagenomic assembler which is able to take advantage of improvements in standard de novo metagenomic assemblers and reference-based assembly, in contrast to recent distributed and parallel assemblers such as Ray Meta~\cite{raymeta}. 

The paper is organized as follows. In Section~\ref{sec:algorithm}, we provide a step-by-step description of the \emph{MetaPar} algorithm. In Section~\ref{sec:example}, we demonstrate the performance of the method on synthetic Illumina sequencer data, using a randomly selected set of $15$ bacterial organisms. We also compute and list the effective coverage and gaps in \emph{MetaPar} alignments to the identified species' genomes. 

\section{An Algorithmic Solution for Parallel Metagenomic Assembly} \label{sec:algorithm}

The following terminology is used throughout the paper. When describing a living organism, we will refer to several taxonomy levels, listed from most general to most specific: life, domain, kingdom, phylum, class, order, family, genus, and species. Each organism has a \emph{genome}, which is a sequence of \emph{bases} over a four letter alphabet. A \emph{read} is a \emph{substring} of a genome or a chromosom (a part of a genome), generated through some sequencing system. The \emph{coverage} of a base in a genome equals the number of reads that contain the base. \emph{Assembly} refers to the process of overlapping reads -- suffix to prefix -- in 
order to reconstruct the original sequence from which the reads came from, or in order to reconstruct sufficiently long substrings of the genome, termed \emph{contigs}. \emph{Alignment} refers to mapping reads onto a given genome. 

The metagenome assembly problem may be formulated as follows: given a mixture of reads from genomes of different species providing sufficiently high coverage, reconstruct the original genomes as accurately as possible via some computationally plausible assembly method. Many different assembly methods for metagenomic data were developed in the past few years, using greedy algorithms, reference-based approaches, algorithms based on deBruijn graphs and Eulerian path searches, and many other techniques~\cite{M2007}. Although the accuracy of the aforementioned assembly methods is high for small metagenomic samples, it quickly deteriorates when the metagenomic data contains fragments of a large number of species. Even more important is the fact that the complexity of most assembly methods grows exponentially with the number of species, leading to poor performance scaling with sample size.
As a result, large metagenomic samples require powerful computers for assembly. This raises the natural question of parallelizing the assembly process.

\begin{figure*}[t]
\begin{centering}
\centerline{\includegraphics[width=.7\textwidth]{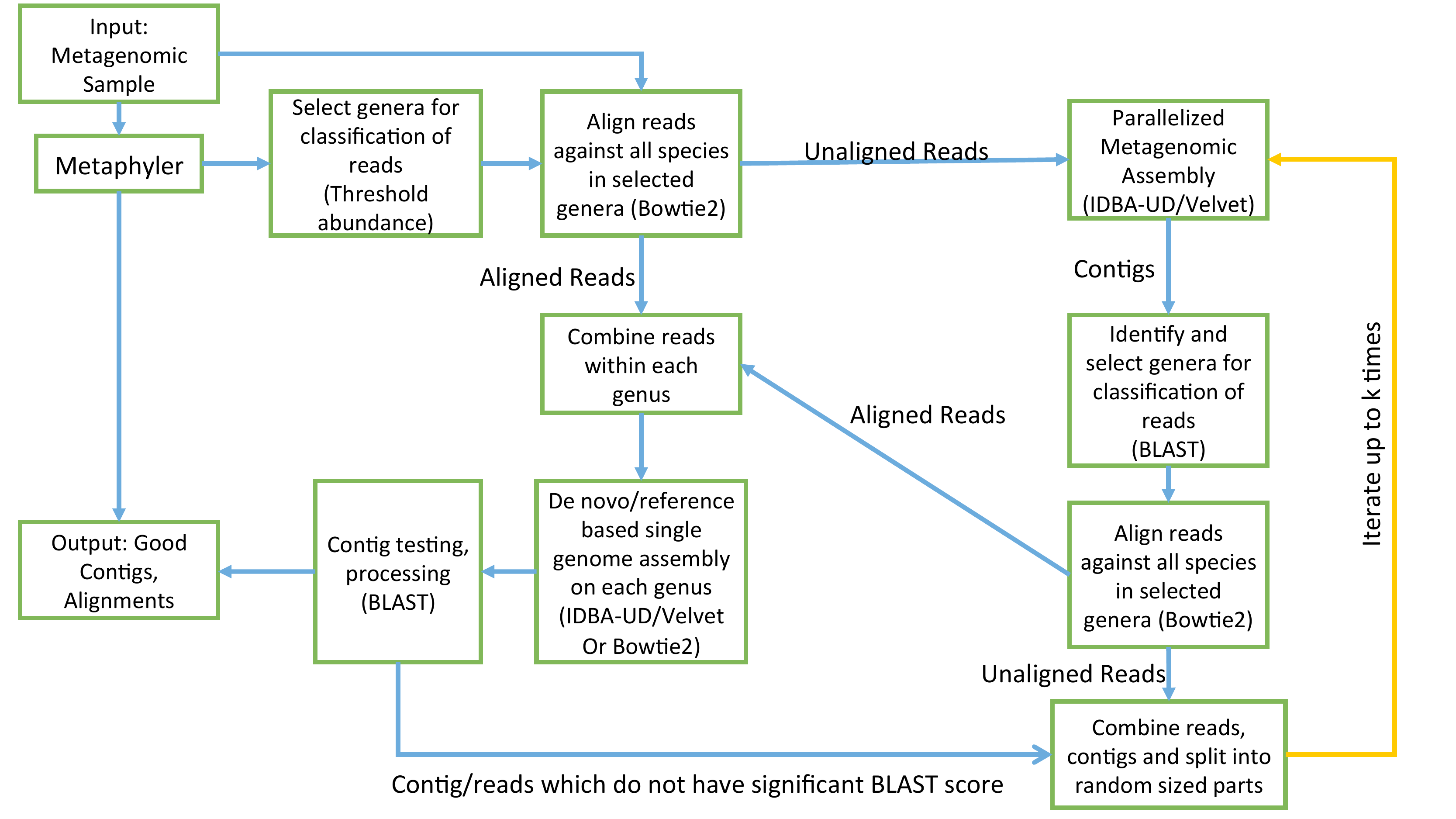}}
\caption{Block Diagram of the \emph{MetaPar} Algorithm for Metagenomic Sequence Assembly.}
\label{fig:MCUIUC}
\end{centering}
\end{figure*}

We next outline the parallel \emph{MetaPar} assembly algorithm that allows for assembling metagenomes containing several hundred species suitable for commodity computers with 16-48+ GB RAM. The running time of all components, aside from calls to standard assemblers, grows linearly with the size of the metagenomic sample. The block diagram of the algorithm is depicted in Figure~\ref{fig:MCUIUC}. Note that some steps in the algorithm are implemented only if the metasample (metagenomic sample) is very large, since in that case, direct assembly is computationally infeasible. The proposed algorithm uses certain techniques related to the authors' MCUIUC algorithm for compression, described in~\cite{mcuiuc}.

\begin{itemize}
\item \textbf{Step 1 (Level I Identification):} The first step of the iterative procedure is to remove as many reads that can be associated with known species from the original sample before running the assembly process. Such a filtering procedure is expected to produce significantly reduced metasamples for subsequent assembly. Filtering is achieved by passing the metasample through a taxonomic identifier, such as \emph{Metaphyler}~\cite{Liuetal2010}\footnote{Very recently, a new approach to species identification was described in~\cite{francis2013} that may outperform \emph{MetaPhyler}. Given that comparing identification software packages is beyond the scope of the paper, we only report results for the more commonly used \emph{MetaPhyler} package.}. The gist of the approach in~\cite{Liuetal2010} is that almost every genomic substring of length exceeding $20$ is unique to a species or a genus. \emph{MetaPhyler} scans through the reads to identify such substrings and links them to a sequenced species. Identification usually amounts to specifying the genera of organisms, and the abundances of the marker sequences. Due to identification errors, the output of the taxonomic classifier contains both false-positive and false-negative results. Selection of the identified genera used for read removal can depend on factors such as the number of identified organisms, genome lengths and identifier abundance. Simulation on synthetic data involving $15$, $30$ and $\geq 60$ species was used to determine simple abundance thresholds, as described in the next section.

\item \textbf{Step 2 (Level I Partitioning of the Dataset):} \begin{enumerate}
\item  Once a group of genera of interest is identified in Step 1, representative reference genomes for alignment of reads are selected. Selection is accomplished by using complete genomes of all species within the identified genera in an alignment procedure performed via \emph{Bowtie2}~\cite{LS2012}. \emph{Bowtie2} is designed for ultra-fast alignment of short reads to long genomes, and its complexity scales roughly linearly with the length of the genomes and number of reads (\emph{Bowtie2}, along with BLAST~\cite{A1990}, is one of the most frequently used alignment algorithms). Given that not all correct genera may have been found by 
\emph{Metaphyler}, some reads will be reported as unaligned. Unaligned in this context refers to not having sufficient sequence similarity to any substrings of the chosen references genomes. The percentage of unaligned reads heavily depends on the size of the metagenome, the number of species involved, as well as the number of identifiers of the species used in the identification software.

\item \textbf{Step 3 (Level II Identification):} Given that \emph{MetaPhyler} may miss identifying a large number of species present in the sample, and that consequently \emph{Bowtie2}-based partitioning may leave a large fraction of reads unclassified, additional identification procedures are needed. Two different procedures are employed based on the volume of the unaligned reads. If the size of the unaligned metagenome is relatively small, metagenomic assembly based on \emph{Velvet}, \emph{SOAP deNovo} or \emph{IDBA-UD}~\cite{Pengetal2012} is used. If the unaligned metagenome is prohibitively large to be assembled by existing assemblers, the unaligned read set is partitioned into an appropriate number of \emph{random subsets} (The largest number of subsets we needed to run on any dataset was eight.).The subsets are processed in parallel by independent assemblers that produce contigs, which can be run through BLAST to identify additional reference genomes for alignment with \emph{Bowtie2}. Given that the partitioning of the dataset is random, many reads may appear as stand-alone contigs at the output of the assembler, and are treated as unaligned reads that need to be re-classified.\footnote{As already mentioned, a means for parallelizing \emph{single genome} assembly that shares some of the classification ideas outlined in this step was first reported in~\cite{W2012}.} 
\item  \textbf{Step 4 (Iterative Re-classification):} Reads that remain unaligned after the described three steps are processed iteratively through Step 3, as long as the number of unaligned reads is higher than a certain threshold or until a maximum number of iterations is executed.
\end{enumerate}
\end{itemize}

\section{Working Example} \label{sec:example}
Metagenomic samples have vastly different sizes~\cite{Q2010}, and the exact number of iterations performed in the assembly process, as well as the exact number of parallel read classes used depends on the metagenomic file size. For example, for the ``CO182: Coal cuttings from Coal bed Methane well site'' sample~\cite{coaldata}, one has to run eight or more instances of a conventional assembler such as IDBA-UD -- in parallel on several machines, or sequentially on one machine -- with about 5.3 GB of reads per assembler. The algorithms were executed on computers equipped with dual Intel Xeon E5630 processors (16 threads) and 48 GB RAM in order to not exhaust the available memory at the 20 kmer level for the modified deBruijn graph search. This problem is further exacerbated with sequencing technologies that produce ``long reads'' ($>$128 bases in the case of IDBA-UD) which require additional data to be maintained for reads during assembly. In many cases, it is not feasible to increase available memory beyond a certain point per machine. Each part required $\sim$140 CPU hours to assemble, which while relatively low is not the critical constraint due to available memory. On the other hand, for most synthetic metagenomes including roughly $15$ species, only one assembler is needed. All computations other than the running of parallel assemblers were performed on a computer equipped with an Intel Core i5 3470 and 16 GB of memory, and were primarily I/O and CPU limited rather than memory limited.  In the former case (CO182), \emph{Metaphyler} only identified genera accountable for $40\%$ of the metareads, while it identified more than $70\%$ of reads in the small synthetic samples. Due to space limitations, we illustrate the performance and the steps of the MetaPar algorithm on a small synthetic sample involving $15$ species, and defer the analysis of real metagenomic samples to the full version of the paper.
 
 \begin{table}[t!]
\caption{A randomly selected set of $15$ species used to illustrate the operating principles of \emph{MetaPar}.} 
\begin{center}
\begin{tabular}{|c|c|c|} \hline
\textbf{$\#$} &	\textbf{Genus}	& \textbf{Species} \\ \hline
1 &	Acidilobus		& Acidilobus\_saccharovorans\_345\_15\_uid51395 \\ \hline
2 &	Alcanivorax 		& Alcanivorax\_borkumensis\_SK2\_uid58169\\ \hline
3 &	Bacteroides	 	& Bacteroides\_fragilis\_YCH46\_uid58195 \\ \hline
4 &	Borrelia		& Borrelia\_garinii\_NMJW1\_uid177081\\ \hline
5 &	Corynebacterium	& Corynebacterium\_pseudotuberculosis\_I19\_uid159673\\ \hline
6 &	Enterobacter		& Enterobacter\_asburiae\_LF7a\_uid72793\\ \hline
7 &	Frankia		& Frankia\_CcI3\_uid58397\\ \hline
8 &	Halomicrobium	& Halomicrobium\_mukohataei\_DSM\_12286\_uid59107\\ \hline
9 &	Helicobacter		& Helicobacter\_pylori\_Shi417\_uid162205\\ \hline
10 &	Lactobacillus		& Lactobacillus\_amylovorus\_GRL1118\_uid160233\\ \hline
11 &	Mobiluncus		& Mobiluncus\_curtisii\_ATCC\_43063\_uid49695\\ \hline
12 &	Mycoplasma		& Mycoplasma\_arthritidis\_158L3\_1\_uid58005\\ \hline
13 &	Odoribacter		& Odoribacter\_splanchnicus\_DSM20712\_uid63397\\ \hline
14 &	Prevotella		& Prevotella\_denticola\_F0289\_uid65091\\ \hline
15 &	Psychroflexus	& Psychroflexus\_torquis\_ATCC700755\_uid54205\\ \hline 
\end{tabular}
\end{center}
\label{speciesList}
\vspace{-20pt}
\end{table}%

\subsection{Simulating the Metagenomic Sample}
Species were randomly selected from the NCBI microbial genome database available at~\cite{wp}. A selected group of $15$ organisms is listed in Table~\ref{speciesList}.  Of the chosen species, \emph{Frankia} has the longest genome with $5,511,253$ bps (base pairs), while \emph{Mycoplasma arthritidis} has the shortest genome with $832,175$ bps.  For each species, we selected the FASTA file containing the complete genome and generated paired reads using the sim\_reads tool accompanying IDBA-UD with settings as in \cite{Pengetal2012}, and coverage depth $100$. The reads from each species were combined to simulate a metagenomic sample from an Illumina sequencer without quality score information (nevertheless, the algorithm can be easily adapted to include quality information as well, and reads stored in the FASTQ format). The resulting metagenomic sample size was $5$ GB in the FASTA format. Note that the chosen synthetic sample, unlike real metagenomic data, had no species from the same genus. How the performance of the algorithm changes in the presence of multiple species per genera will be described in the full version of the paper.

\subsection{Step 1: Metaphyler Genus Identification}
\begin{figure}[!t]
\centering
\includegraphics[width=3.0in]{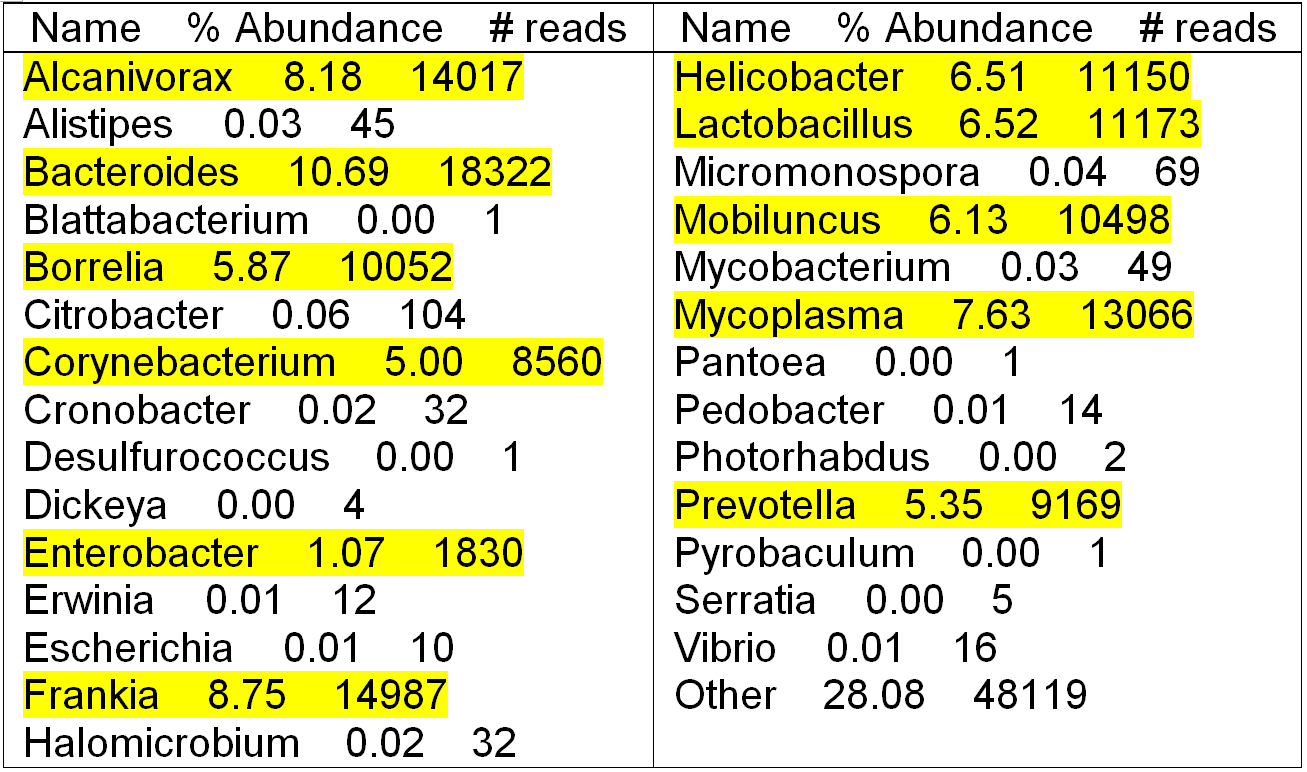}
\caption{Output of Metaphyler on randomly selected set of organisms, shown in Table ~\ref{speciesList}. 11 genera with abundance higher than 1$\%$ are highlighted.}
\label{fig:metaphyler-output}
\end{figure}

\emph{MetaPhyler} takes the simulated metagenomic reads as inputs and outputs their taxonomy classifications. We focused our attention on genus classification, as it is the finest reliable level provided by \emph{MetaPhyler}.  The \emph{MetaPhyler} genus-level output for the input metagenomic reads is given in Fig.~\ref{fig:metaphyler-output}. Note that each genus identifier appears in the table with the number of reads containing markers and the abundance level of such reads. We only selected genera with abundance or number of reads exceeding a certain threshold. The choice of the threshold is governed by many parameters, including the number of estimated organisms, their genome lengths, the number of known markers in the genomes, as well as the actual output of \emph{MetaPhyler}. As a guideline, we used the threshold criteria listed in Table ~\ref{threshold}. Of the $28$ genera identified, $11$ satisfied the threshold criteria, which in this case amounted to more than $1\%$ abundance or at least $1000$ reads. The selected set of $11$ genera contains \emph{all} true positives and \emph{no} false positives.  However, \emph{MetaPhyler} missed identifying the genera of four organisms present in the metagenomic sample, namely \emph{Acidilobus}, \emph{Halomicrobium}, \emph{Odoribacter}, and \emph{Psychroflexus}.  

\begin{table}[t!]
\caption{A threshold criteria to select top genera of \emph{MetaPhyler} output.} 
\begin{center}
\begin{tabular}{|c|c|c|} \hline
\textbf{$\#$ of reported species by Metaphyler}  	&	\textbf{Abundance ($\%$)}	& \textbf{$\#$ Reads} \\ \hline
$\leq$ 35 					  		&	1					& $\geq 1000$ \\ \hline
$>35, \leq 60$						&	0.3 					& $\geq 300$ \\ \hline
$>60$ 							&	N/A	 				& $>100$ \\ \hline
\end{tabular}
\end{center}
\label{threshold}
\end{table}

Species within the metagenomic mixture were identified through an additional procedure described in the next subsection.

\subsection{Step 2: Read Classification}
For the purpose of classifying the reads based on similarity to the selected reference genomes, we used the \emph{Bowtie2} algorithm~\cite{LS2012}. Using all species of the $11$ chosen genera above and building a \emph{Bowtie2} index provided a very good metagenomic read alignment rate, equal to $70.29\%$.  Approximately $30\%$ of the reads were not aligned to any reference genomes, so these unaligned reads were assembled via IDBA-UD. The longest 30  resulting contigs were passed through BLAST.  BLAST identified 15 of the contigs as \emph{Odoribacter splanchnicus}, 8 as \emph{Acidilobus saccharovorans}, 6 as \emph{Psychroflexus torquis}, and 1 as \emph{Halomicrobium mukohataei}, which were exactly the four species missed by \emph{MetaPhyler} in Step 1. These 4 species and the 25 species listed in table ~\ref{Topspecies} were used as reference genomes for the second iteration of \emph{Bowtie2}, and the read alignment rate was $99.94\%$.

\begin{table}[t!]
\caption{Up to 9 biggest species files per genus assigned by Bowtie after the first iteration.} 
\begin{center}
\begin{tabular}{|>{\centering\arraybackslash}p{0.1cm}|>{\centering\arraybackslash}p{1.5cm}|>{\centering\arraybackslash}p{6.2cm}|} 	\hline
\textbf{$\#$} 	&	\textbf{Genus}	& \textbf{Accession $\#$ and Description} 								\\ \hline
1			&	Alcanivorax 		& NC\_008260 (Alcanivorax borkumensis SK2)								\\ \hline
2 			&	Bacteroides	 	& NC\_006347 (Bacteroides fragilis YCH46) \newline
							   NC\_016776 (Bacteroides fragilis 638R)  \newline
							   NC\_003228 (Bacteroides fragilis NCTC 9343)							\\ \hline
3 			&	Borrelia		& NC\_018747 (Borrelia garinii NMJW1)	\newline			
							   NC\_017717 (Borrelia garinii BgVir) \newline
							   NC\_006156 (Borrelia garinii PBi )									\\ \hline
4 			&	Corynebacterium	& NC\_017303 (Corynebacterium pseudotuberculosis I19) \newline						 
							   NC\_017031 (Corynebacterium pseudotuberculosis P54B96) \newline
							   NC\_017462 (Corynebacterium pseudotuberculosis 267) \newline
							   NC\_017306 (Corynebacterium pseudotuberculosis 42/02-A) \newline
							   NC\_017305 (Corynebacterium pseudotuberculosis PAT10) \newline
							   NC\_017301 (Corynebacterium pseudotuberculosis C231) \newline
							   NC\_017300 (Corynebacterium pseudotuberculosis 1002) \newline
							   NC\_016781 (Corynebacterium pseudotuberculosis 3/99-5) \newline
							   NC\_014329 (Corynebacterium pseudotuberculosis FRC41)					\\ \hline
5 			&	Enterobacter		& NC\_015968 (Enterobacter asburiae LF7a)								\\ \hline
6			&	Frankia		& NC\_007777 (Frankia sp. CcI3)			 							\\ \hline
7 			&	Helicobacter		& NC\_017739 (Helicobacter pylori Shi417)								\\ \hline
8 			&	Lactobacillus		& NC\_017470 (Lactobacillus amylovorus GRL1118) \newline
							   NC\_015214 (Lactobacillus acidophilus 30SC) \newline
							   NC\_014724 (Lactobacillus amylovorus GRL 1112)							\\ \hline
9 			&	Mobiluncus		& NC\_014246 (Mobiluncus curtisii ATCC 43063)							\\ \hline
10 			&	Mycoplasma		& NC\_011025 (Mycoplasma arthritidis 158L3-1)							\\ \hline
11 			&	Prevotella		& NC\_015311 (Prevotella denticola F0289)								\\ \hline
\end{tabular}
\end{center}
\label{Topspecies}
\vspace{-10pt}
\end{table}%

\subsection{Assembler performance evaluation}

One of the most commonly used statistics for assessing the performance of an assembler is the N50 statistic on contig lengths. The N50 statistic is a threshold value for the length, such that contigs of length longer than or equal to the threshold account for roughly $50\%$ of the total contig length found by the assembler. In other words, it is helpful to think of the N50 parameter as the median of the contig length distribution. Since multiple lengths may satisfy this criteria, the N50 value is often chosen to be the average of all thresholds that satisfy the terms of the definition.

The utility of the N50 value for assessing assembler performance is questionable, since it does not convey important information about what percentage of the length of underlying genomes is actually covered by the contigs and to what extent. This is especially true for reference based assembly. To mitigate this problem, we introduced two performance measures, termed the \emph{effective average coverage} and the \emph{effective gap}. The effective coverage measures the average number of times a base in the genome is covered by the longest matches in each read aligned via \emph{Bowtie2} without errors. Similarly, the effective gap measures the number of bases that were not covered by the longest matches contained in each read aligned via \emph{Bowtie2}. 
\begin{table}[t!]
\caption{Table of Effective Coverage \& Gap for Reference-Based Assembly with Bowtie2} 
\begin{center}
\begin{tabular}{|l|l|l|}
\hline Genus	&Effective Coverage	&Effective Gap	\\ \hline
Acidilobus	&99.866	&5.5	\\ \hline
Alcanivorax	&99.863	&2	\\ \hline
Bacteroides	&99.947	&0	\\ \hline
Borrelia	&99.863	&3	\\ \hline
Corynebacterium	&99.867	&2	\\ \hline
Enterobacter	&99.867	&1	\\ \hline
Frankia	&99.869	&1	\\ \hline
Halomicrobium	&99.866	&1	\\ \hline
Helicobacter	&99.867	&4	\\ \hline
Lactobacillus	&99.866	&4	\\ \hline
Mobiluncus	&99.867	&4	\\ \hline
Mycoplasma	&99.865	&0	\\ \hline
Odioribacter	&99.751	&0	\\ \hline
Prevotella	&99.861	&2	\\ \hline
Psychroflexus	&99.856	&1	\\ \hline
\end{tabular}
\end{center}
\label{egapecovtable}
\vspace{-20pt}
\end{table}%
The effective gap and effective coverage for the given example are listed in Table \ref{egapecovtable}. 

As can be seen from the table, the effective coverage is very large, exceeding $99.751\%$ for all metasample organisms. The observed gaps are extremely small, with the worst performance observed for the genera \emph{Acidilobus}.
A quick look at Table III reveals that the selected organism in this genus was not identified by \emph{Metaphyler} and may have consequently had many substrings shared by other organisms. This may be a plausible explanation for read misclassification and consequently high effective gap. 

\section*{Acknowledgment}
This work was supported in part by NSF grants CCF 0809895, CCF 1218764, Emerging Frontiers for Science of Information Center, CCF 0939370 and U.S. Defense Threat Reduction Agency through subcontract 147755 at the University of Illinois from prime award HDTRA1-10-1-0086.
The authors also gratefully acknowledge many useful discussions with Prof. Jian Ma and Xiaolong Wu at the University of Illinois, Urbana-Champaign.

\section{Conclusions}
We have described a new parallelizable framework for metagenomic assembly in which the computational time for most steps grow linearly in time with the size of the metagenomic sample. The algorithm identifies genera and species in order to use reference based assembly to reduce the amount of standard de novo assembly required. Performance was illustrated on a synthetic sample of 15 species. Further work includes designing schemes for efficient partitioning of reads akin to TIGER, incorporation of phylogenic aligners and incorporation of other classifiers for reads.

\end{document}